\documentclass[twocolumn]{aastex6}

\begin{document}

\title{PHOTOSPHERIC VELOCITY STRUCTURES DURING THE EMERGENCE OF SMALL ACTIVE REGIONS ON THE SUN}

\author{Anna Khlystova\altaffilmark{1}}
\affil{Institute of Solar-Terrestrial Physics SB RAS \\
Lermontov St., 126a \\
664033 Irkutsk, Russia}

\and

\author{Shin Toriumi\altaffilmark{2}}
\affil{National Astronomical Observatory of Japan \\
2-21-1 Osawa, Mitaka \\ 
Tokyo 181-8588, Japan}

\altaffiltext{1}{hlystova@iszf.irk.ru}
\altaffiltext{2}{shin.toriumi@nao.ac.jp}

\begin{abstract}

We study the plasma flows in the solar photosphere during the emergence of two 
small active regions, NOAA 9021 and 10768. Using \textit{Solar and Heliospheric 
Observatory/}Michelson Doppler Imager data, we find that the strong plasma 
upflows appear at the initial stage of active region formation, with maximum 
upflow velocities of $-$1650\,m\,s$^{-1}$ and $-$1320\,m\,s$^{-1}$. The 
structures with enhanced upflows have size $\sim$8\,Mm in diameter, and they 
exist for 1\,--\,2 hr. The parameters of the enhanced upflows are consistent 
with those of the large active region NOAA 10488, which may suggest the 
possibility that the elementary emerging magnetic loops that appear at the 
earliest phase of active region formation have similar properties, irrespective 
of scales of active regions. Comparison between the observations and a numerical 
simulation of magnetic flux emergence shows a striking consistency. We find that 
the driving force of the plasma upflow is at first the gas pressure gradient and 
later the magnetic pressure gradient.

\end{abstract}

\keywords{Sun: photospheres --- Sun: magnetic fields --- Sun: activity}

\section{Introduction}

Magnetic fields on the Sun are thought to be generated in the convective zone. 
They emerge to the surface as $\Omega$-loops under the action of magnetic 
buoyancy and convective upflows, eventually forming active regions at the solar 
surface (\citealt{parker55}; see reviews by \citealt{fan09} and 
\citealt{cheung14}). Active regions are different in magnetic flux content 
\citep{garciadelarosa84,zwaan87}. In small active regions, magnetic flux in each 
polarity is $10^{20}$\,Mx\,–-\,0.5$\times10^{21}$\,Mx; in white light, they 
contain only pores. In large active regions, the magnetic flux of each polarity 
exceeds 0.5$\times10^{21}$\,Mx; in white light, pores and sunspots are observed 
where the magnetic field is concentrated.

From the viewpoint of the active region development, the photospheric plasma 
flow at the initial stage of magnetic flux emergence is one of the most 
fascinating topics. The continuous full-disk observations of the Sun by 
\textit{Solar and Heliospheric Observatory (SOHO)/}Michelson Doppler Imager 
(MDI) and \textit{Solar Dynamics Observatory (SDO)/}Helioseismic and Magnetic 
Imager (HMI) have allowed us to realize this task. For example, 
\cite{grigorev07} found enhanced plasma upflow in the photosphere, with velocity 
up to 1.7\,km\,s$^{-1}$ at the beginning of emergence of the large active region 
NOAA 10488. Later, \cite{grigorev09} studied the initial stage of NOAA 10488 
formation in detail. On the basis of a statistical analysis, \cite{khlystova11} 
obtained a center-to-limb dependence of the negative Doppler velocities, which 
shows that at the initial stage of active region formation in the photosphere, 
the horizontal outflow plasma velocities are higher than the vertical upflows 
ones. \cite{toriumi12,toriumi14} detected a strong horizontal divergent flow 
that starts about 1 hr before the beginning of the magnetic flux emergence in 
the photosphere. \cite{khlystova13a} found the strong horizontal divergent flows 
of photospheric plasma during the first hours of active region formation with 
high magnetic flux growth rate. \cite{khlystova13b} carried out the statistical 
analysis on the relation of the Doppler velocities of vertical and horizontal 
plasma flows, with some parameters of the magnetic fields at the initial stage 
of active region formation.

Subsurface plasma flows in the emerging flux regions have been intensively 
investigated using local helioseismology. Subphotospheric signatures of the 
magnetic flux emergence before the active region appearance on the solar surface 
are surveyed using helioseismic holography 
\citep{braun95,hartlep11,birch13,barnes14}, ring-diagram analysis 
\citep{komm08}, and time-distance helioseismology 
\citep{ilonidis11,kholikov13,ilonidis13,kosovichev16,singh16}. In some studies, 
subphotospheric velocities of the magnetic flux emergence in large active 
regions were estimated. \cite{kosovichev00} have determined that the emerging 
magnetic flux of the active region NOAA 8131 crosses the top 10\,Mm of the 
convective zone at about 1.3\,km\,s$^{-1}$. By tracking the sound-speed 
perturbations, \cite{zharkov08} estimated the velocity of magnetic flux 
emergence through the top 21.7\,Mm of the convective zone in the active region 
NOAA 10790 to be $\sim$1\,km\,s$^{-1}$. \cite{ilonidis11} obtained the 
0.3\,--\,0.6\,km\,s$^{-1}$ average velocity of the magnetic flux emergence in 
four large active regions from the 60\,Mm depth to the surface. Analysis by 
\cite{toriumi13b} suggested a gradual deceleration of the magnetic flux in the 
subphotospheric layers of NOAA 10488. According to their calculations, the 
velocity of the magnetic flux emergence is several km\,s$^{-1}$ at the depth of 
10\,--\,15\,Mm, $\sim$1.5\,km\,s$^{-1}$ at 5\,--\,10\,Mm, and 
$\sim$0.5\,km\,s$^{-1}$ at 2\,--\,5\,Mm. \cite{ilonidis13} estimated the rise 
velocity of the magnetic flux in NOAA 10488 at 42\,--\,75\,Mm depth to be 
$\sim$1\,km\,s$^{-1}$. \cite{kosovichev16} investigated the subsurface dynamics 
of the emerging active region NOAA 11726, the largest region during the first 5 
year observations of the \textit{SDO/}HMI. They found that the magnetic flux in 
deep (42\,--\,75\,Mm) and subphotospheric (5\,--\,20\,Mm) layers emerges at the 
same velocity of $\sim$1.4\,km\,s$^{-1}$. On the contrary, \cite{birch16} 
obtained a low upper limit of the subsurface velocity of the magnetic flux rise. 
In their research, the rising velocity of the magnetic fields at the 20 Mm depth 
does not exceed 150\,m\,s$^{-1}$.

Another possibility for probing the emerging magnetic flux in the subsurface 
region is to directly compare the observation of the flux emergence with 
simulations. Since the initial appearance of magnetic flux and the accompanying 
flow may be largely affected by the rising flux in the interior, a comparative 
study of observation and simulation may provide clues for understanding the 
physical state of the subsurface magnetic flux.

In this paper, we focus on flow structures in the photosphere during the initial 
appearance of small active regions. Section \ref{sec:obs} describes the data 
that we use and their processing. Section \ref{sec:ar} presents the analysis 
results of the flow structures in the emerging active regions. We compare the 
observational results with a numerical simulation in Section \ref{sec:phys}. 
Discussion and conclusions are shown in Sections \ref{sec:disc} and 
\ref{sec:conc}, respectively.

\section{Observations and data processing} \label{sec:obs}

We studied plasma flows in the photosphere during the emergence of two small 
active regions NOAA 9021 and 10768. The observed flows were compared with that 
of the large active region NOAA 10488 \citep{grigorev07}. Table 
\ref{tab:ar_main1} shows the time of emergence and position on the solar disk of 
the three active regions. These active regions are sub-equatorial and are 
located at an approximately equal heliocentric distance from the solar center, 
which allows us to compare them without taking into account the projection 
effect.

\floattable
\begin{deluxetable}{ccccc}
\tablecaption{Time, coordinates, and heliocentric angle $\theta$ at the beginning of emergence of the active regions \label{tab:ar_main1}}
\tablecolumns{5}
\tablenum{1}
\tablewidth{0pt}
\tablehead{\colhead{Active Region} & \colhead{Time of Emergence} & \colhead{Coordinates} & \colhead{$\theta$}}
\startdata
 9021 & 2000.05.27 $<$16:24 UT & N03 E35 B$_{0}-$1.1 & 36$^{\circ}$ \\
10768 & 2005.05.25 -- 17:42 UT  & S08 W32 B$_{0}-$1.3 & 33$^{\circ}$ \\
\hline
10488 & 2003.10.26 -- 09:07 UT  & N08 E30 B$_{0}+$4.9 & 30$^{\circ}$ \\
\enddata
\end{deluxetable}

We analyzed observational data obtained by the MDI onboard the \textit{SOHO} 
\citep{scherrer95}. Between 1996 and 2011, \textit{SOHO/}MDI recorded the line 
of sight magnetic field strength, Doppler velocity, and continuum intensity of 
the full solar disk in the photospheric line Ni\,{\sc i} 6768\,\AA. The temporal 
resolution of the magnetogram and Dopplergram is 1 minute; continuum image, 96 
minutes. Due to the high temporal resolution of the magnetogram and Dopplergram, 
it is possible to track changes in the magnetic fields and plasma flows in great 
detail from the earliest phase of active region formation. The spatial 
resolution of the data is 4$''$, with a 2$''$ pixel size. Due to the low spatial 
resolution, the data do not contain a contribution of high velocities of 
granular convection.

We made cutout data of the emerging magnetic flux regions with a size of 
160$''\times$140$''$ (or 80$\times$70 pixels) from the time series of full solar 
disk images, while taking into account solar rotation. The displacement of the 
region under investigation was calculated based on the differential rotation law 
for photospheric magnetic fields \citep{snodgrass83} and cross-correlation 
analysis between magnetograms. To obtain better coalignments between the data, 
we picked up identical fragments from the magnetogram, Dopplergram, and 
continuum images, and superposed them against each other.

The Dopplergram contains velocity components of various effects that are not 
uniformly distributed over the solar disk \citep{howe11}. We removed these 
effects by applying the method proposed by \cite{grigorev07}. Then, we applied a 
moving average of five images to reduce photospheric oscillations. In the 
Dopplergrams, the negative (positive) velocity corresponds to the blue (red) 
shift, or plasma motion toward (away from) the observer.

For each active region, we measured the following parameters: the total area of 
pores and sunspots $S_{\rm spot}$, which is defined as the area inside the 
isoline of 85\% of the average quiet-Sun intensity, taking into account the 
projection effect; the total unsigned magnetic flux $\Phi$, measured within the 
isoline of $\pm$60\,G, taking into account the projection effect and subtracting 
the background magnetic flux that is present before the active region 
appearance; the growth rate of the total unsigned magnetic flux $d\Phi/dt$ and 
the maximum magnetic field strength $H_{\rm max}$ during the first 12 hr of the 
flux emergence; the absolute maximum values of negative and positive Doppler 
velocities $V_{\rm max-}$ and $V_{\rm max+}$, and also the mean negative Doppler 
velocity $V_{\rm mean+}$, during the first 4 hr of the active region 
development; and the maximum area of velocity structure inside the 
$-$500\,m\,s$^{-1}$ isoline $S_{\rm vel}$. The calculation region of these 
parameters was limited to the region of the magnetic flux emergence. To trace 
velocity structures, we used the $-$500\,m\,s$^{-1}$ isoline, because 
photospheric plasma flows, which are not associated with the emerging magnetic 
flux, are generally below this level.

In this paper, we consider small active regions that are taken from our earlier 
statistical study \citep{khlystova13b}, in which we used \textit{SOHO/}MDI data 
to study photospheric plasma flows accompanying the magnetic flux emergence. As 
a future work, we plan to use \textit{SDO/}HMI data, which have better temporal 
and spatial resolutions, for further investigations.

\section{Long-term Evolution of Small Active Regions and Their Initial Magnetic 
Flux Emergence} \label{sec:ar}

\subsection{NOAA 9021}

\textit{Long-term evolution.} Active region NOAA 9021 begins to form on 2000 May 
27, during the data gap between 13:58 and 16:24 UT at N03 E35 (see Table 
\ref{tab:ar_main1}). This active region appears almost in the same area of the 
solar disk as NOAA 10488.

\begin{figure*}
\centerline{\includegraphics[width=0.9\textwidth]{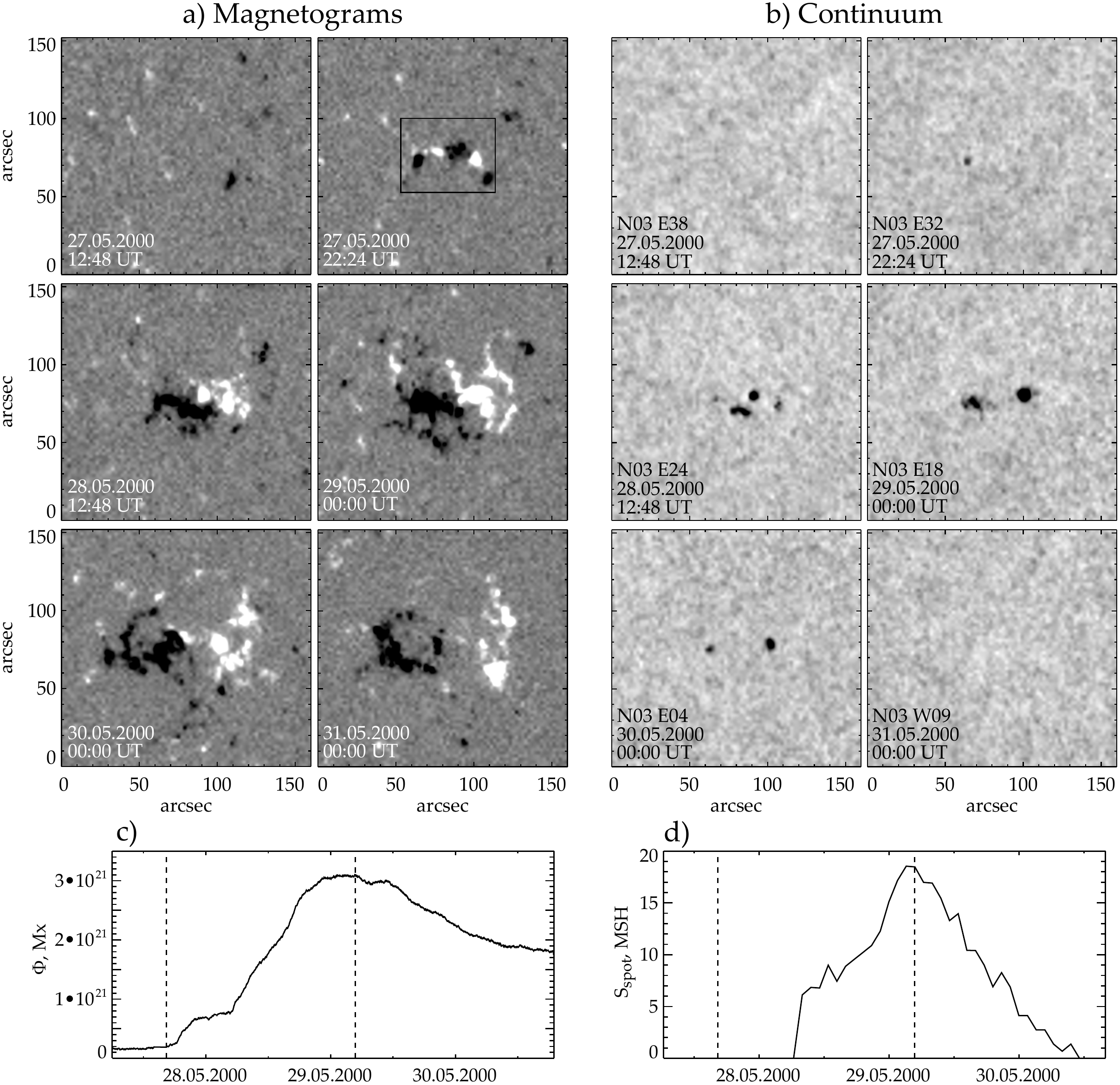}}
\caption{Development of the active region NOAA 9021: (a) the photospheric line 
of sight magnetograms saturated at $\pm$200\,G; (b) the continuum; the time 
variation of (c) the total unsigned magnetic flux; and (d) the sunspot area, 
where vertical dashed lines mark the beginning and maximum of active region 
development. The black box on the magnetogram indicates the field of view of 
Figure \ref{fig:fig2}.}
\label{fig:fig1}
\end{figure*}

NOAA 9021 appears as two emerging bipoles, which later merge together and form a 
single bipolar region (Figure \ref{fig:fig1}\,a). The magnetic axis connecting 
opposite magnetic polarities becomes oriented along the equator 12 hr after the 
beginning of the active region emergence. The average magnetic flux growth rate 
during the first 12 hr of the active region development is 
0.5$\times10^{20}$\,Mx\,hr$^{-1}$, which is about 1/8 of that in NOAA 10488. 
NOAA 9021 reaches its maximal development in approximately 1.5 days with total 
unsigned magnetic flux 2.9$\times10^{21}$\,Mx (Figure \ref{fig:fig1}\,c). Soon 
after that, the active region decays quickly.

It is seen from the continuum images that only pores without penumbrae are 
formed in this active region (Figure \ref{fig:fig1}\,b). Two pores with 
approximately equal sizes are observed in leading and following polarities in 
the maximum of the active region development. Their total area reaches 19 MSH 
(Figure \ref{fig:fig1}\,d).

\begin{figure*}
\centerline{\includegraphics[width=0.7\textwidth]{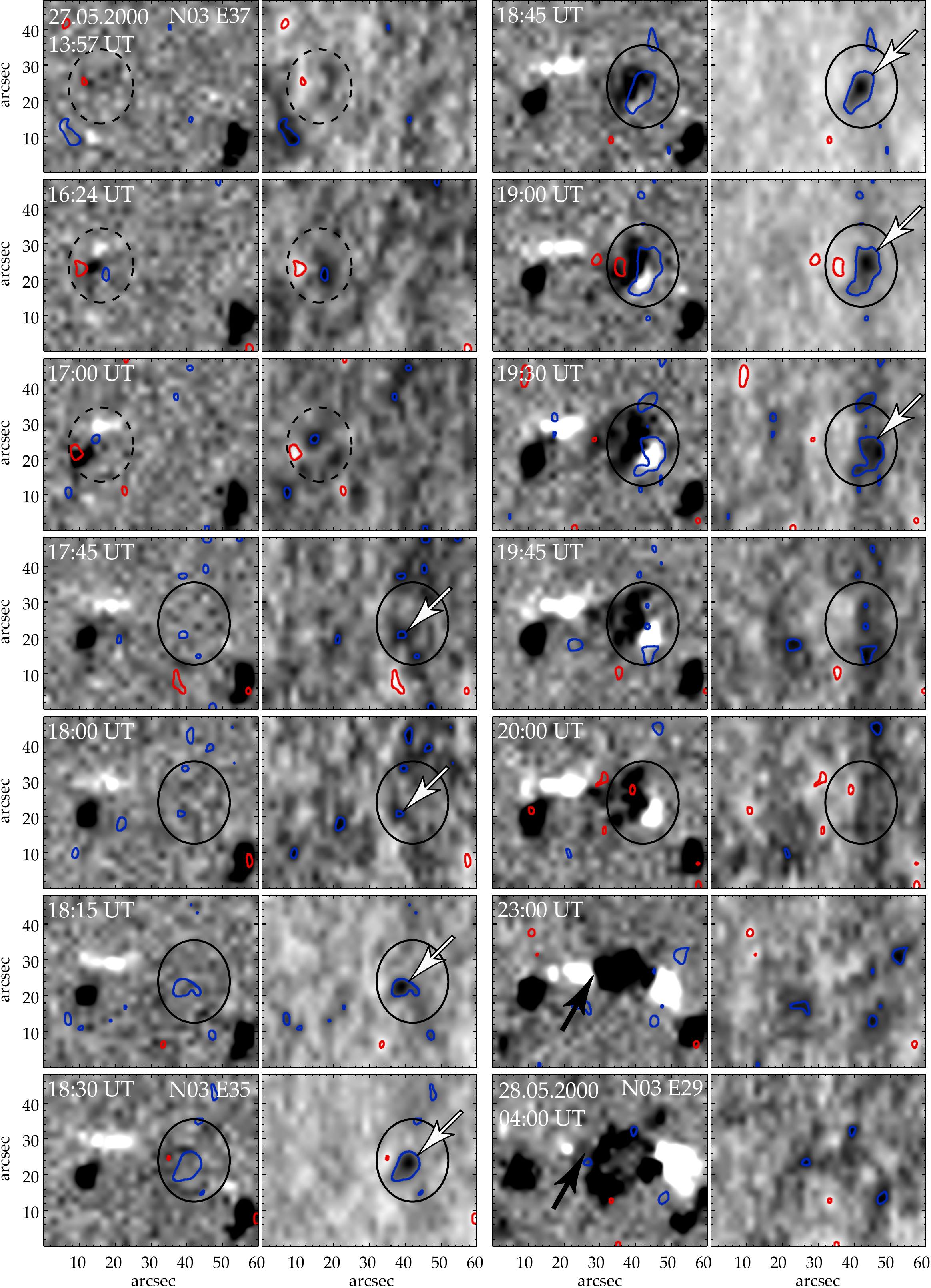}}
\caption{Emergence of the active region NOAA 9021 on 2000 May 27. For each 
moment, the magnetograms saturated at $\pm$100\,G are shown on the left; the 
Dopplergrams are shown on the right (dark color marks negative magnetic fields 
and Doppler velocities; light color marks the positive ones). The isolines of 
velocity $\pm$500\,m\,s$^{-1}$ are superimposed on magnetograms and Dopplergrams 
(the blue isoline represents negative Doppler velocity; the red line represents 
the positive one). The first magnetic flux emergence is marked by a dashed 
ellipse; the second (main) magnetic flux emergence is marked by a solid ellipse. 
See the comments in the text.}
\label{fig:fig2}
\end{figure*}

\textit{Initial stage of magnetic flux emergence.} In NOAA 9021, the first 
bipole starts emergence in the data gap between 13:58 and 16:24 UT, May 27 
(Figure \ref{fig:fig2}, marked by dotted ellipse). Just after the data gap, the 
appeared magnetic polarities have small sizes, and their maximum magnetic field 
strength does not exceed 150\,G. During this magnetic flux emergence, the 
largest negative and positive Doppler velocities do not exceed 
$-$700\,m\,s$^{-1}$ and 800\,m\,s$^{-1}$, respectively.

Approximately 2 hr after the beginning of first magnetic flux emergence, the 
second (main) bipole emerges to the west of the first pair (Figure 
\ref{fig:fig2}, marked by solid ellipse). The negative polarity inside the 
$-$60\,G isoline appears at 18:31 UT, while the positive polarity begins to form 
from 18:48 UT. The lag between the appearance times of both polarities may be 
due to the predominance of horizontal magnetic fields in the earliest phase of 
the magnetic flux emergence. In the later phase, the first and second (i.e., 
main) emerging bipoles show magnetic flux cancellation (Figure \ref{fig:fig2}, 
marked by black arrows), which can be explained as magnetic reconnection between 
the two $\Omega$-shaped loops.

The plasma upflow region, which is delineated by the $-$500\,m\,s$^{-1}$ isoline 
(blue), appears at the location of the second bipole at 17:44 UT (Figure 
\ref{fig:fig2}, marked by white arrows). However, the velocity values here do 
not exceed $-$650\,m\,s$^{-1}$. Enhancement of the negative Doppler velocity 
begins at 18:04 UT, i.e., 27 minutes before the magnetic flux emergence. As the 
photospheric magnetic flux develops, the $-$500\,m\,s$^{-1}$ isoline covers a 
considerable fraction of the magnetic elements. The region of enhanced negative 
Doppler velocity starts to split at 19:32 UT and disappears before 19:50 UT. 
Also, a small region of the positive Doppler velocity, characterized by the 
500\,m\,s$^{-1}$ isoline, is observed in the following negative polarity from 
18:51 UT to 19:10 UT. However, the maximum velocity does not exceed 
720\,m\,s$^{-1}$.

\begin{figure*}
\centerline{\includegraphics[width=0.45\textwidth]{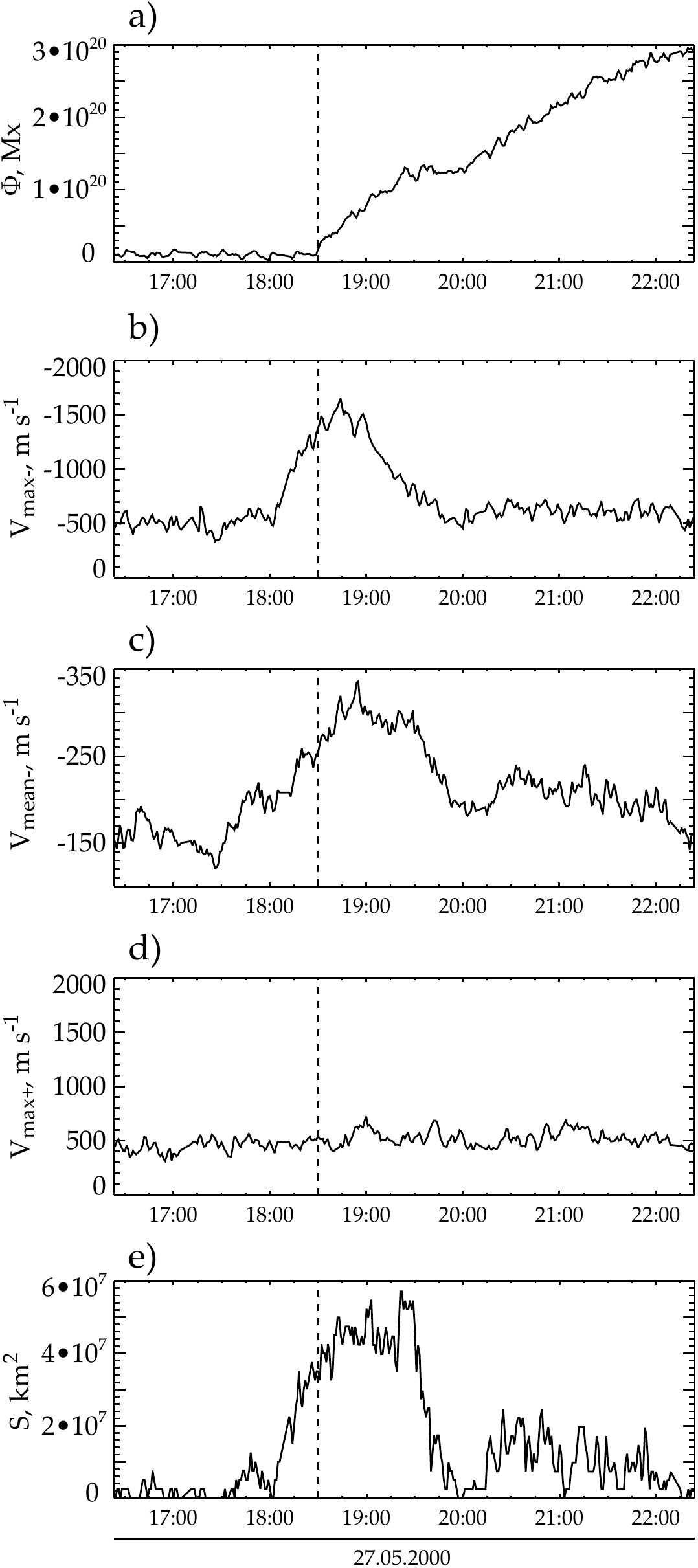}}
\caption{Emergence of the second (main) magnetic flux in the active region NOAA 
9021. Time variation of (a) the total unsigned magnetic flux, (b) the maximum 
negative Doppler velocity, (c) the mean negative Doppler velocity, (d) the 
maximum positive Doppler velocity, and (e) the area of negative Doppler 
velocities inside the $-$500\,m\,s$^{-1}$ isoline. The vertical dashed line 
marks the beginning of magnetic flux emergence.}
\label{fig:fig3}
\end{figure*}

In Figure \ref{fig:fig3}, we plot the temporal evolution of the total magnetic 
flux, the maximum and mean values of negative and positive Doppler velocities, 
and the area of negative Doppler velocities inside the $-$500\,m\,s$^{-1}$ 
isoline. Whereas the magnetic flux emergence begins at 18:31 UT (Figure 
\ref{fig:fig3}\,a), the maximum negative Doppler velocity shows a rapid increase 
27 minutes before, at 18:04 UT (Figure \ref{fig:fig3}\,b). The negative Doppler 
velocity reaches its peak value of $-$1650\,m\,s$^{-1}$ at 18:44 UT, and then 
begins to decrease. The mean negative Doppler velocity also greatly enhances to 
about twice the original value (Figure \ref{fig:fig3}\,c). On the other hand, 
the maximum positive Doppler velocity does not exceed 720\,m\,s$^{-1}$ (Figure 
\ref{fig:fig3}\,d). The area of velocity structure inside the 
$-$500\,m\,s$^{-1}$ isoline lasts for about 2 hr, with its peak value of $\sim 
5.7\times 10^{7}\ {\rm km}^{2}$ (Figure \ref{fig:fig3}\,e).

\subsection{NOAA 10768}

\textit{Long-term evolution.} NOAA 10768 begins to form on 2005 May 25, at 17:42 
UT at S08 W32 (see Table \ref{tab:ar_main1}) -- that is, near the equator, 
approximately at the same heliocentric angle as the large active region NOAA 
10488.

\begin{figure*}
\centerline{\includegraphics[width=0.9\textwidth]{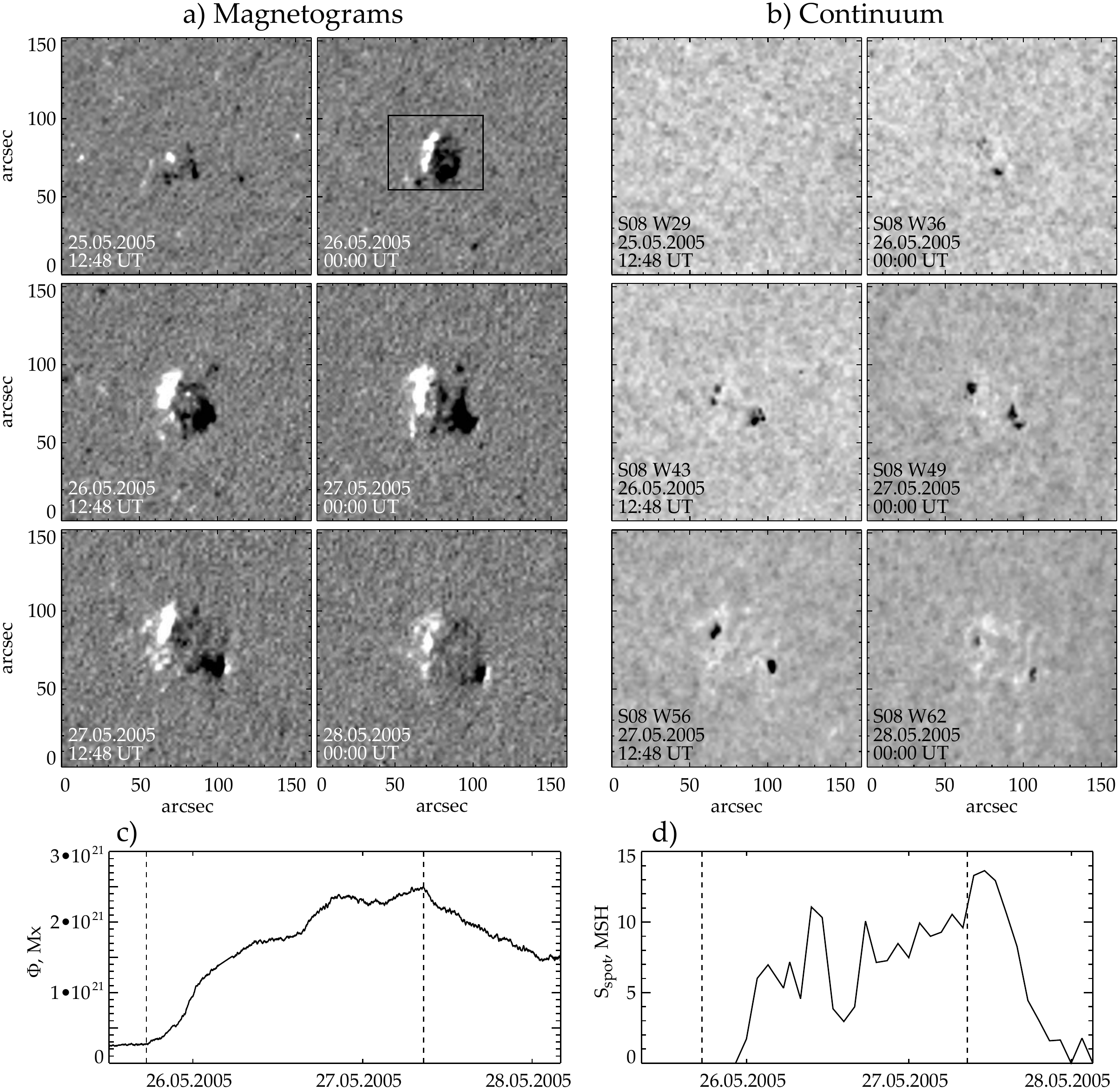}}
\caption{Development of the active region NOAA 10768. Notations are the same as 
in Figure \ref{fig:fig1}. The black box on the magnetogram indicates the 
field of view of Figure \ref{fig:fig5}.}
\label{fig:fig4}
\end{figure*}

The magnetic flux emerges in a narrow region, where a number of small-scale 
magnetic elements of positive and negative polarities appear at the photosphere 
(Figure \ref{fig:fig4}\,a). Although this active region emerges near the 
equator, it shows a large negative tilt angle, which deviates from the Joy's law 
\citep{hale19}. According to the DPD 
catalog\footnote{\url{http://fenyi.solarobs.unideb.hu/DPD/index.html}}, the tilt 
angle in a maximum of the active region development is $-24^{\circ}$.

The average magnetic flux growth rate during the first 12 hr of the active 
region appearance is $\sim$1$\times10^{20}$\,Mx\,hr$^{-1}$, which is about 1/4 
of that in NOAA 10488. Time variation of the total magnetic flux shows several 
points of inflection (Figure \ref{fig:fig4}\,c), indicating that the development 
of this active region is caused by episodes of magnetic flux emergence. The 
active region reaches its maximal development in about 1.5 days, with the total 
unsigned magnetic flux being 2.2$\times10^{21}$\,Mx.

The continuum images show the appearance of pores in this active region (Figure 
\ref{fig:fig4}\,b). Similar to NOAA 9021, at its maximum of the NOAA 10768 
development, there are two pores of an approximately equal size in leading and 
following polarities. The largest total area of the pores reaches 14 MSH (Figure 
\ref{fig:fig4}\,d). On May 29, about 3.5 days after its initial appearance, the 
active region rotates over the west limb. Judging from the temporal evolution 
(Figures \ref{fig:fig4}\,c and d), it is likely that this active region reaches 
its maximum on the near side of the solar disk. Also indicative is that NOAA 
10768 completely decayed before the next solar rotation.

\begin{figure*}
\centerline{\includegraphics[width=0.7\textwidth]{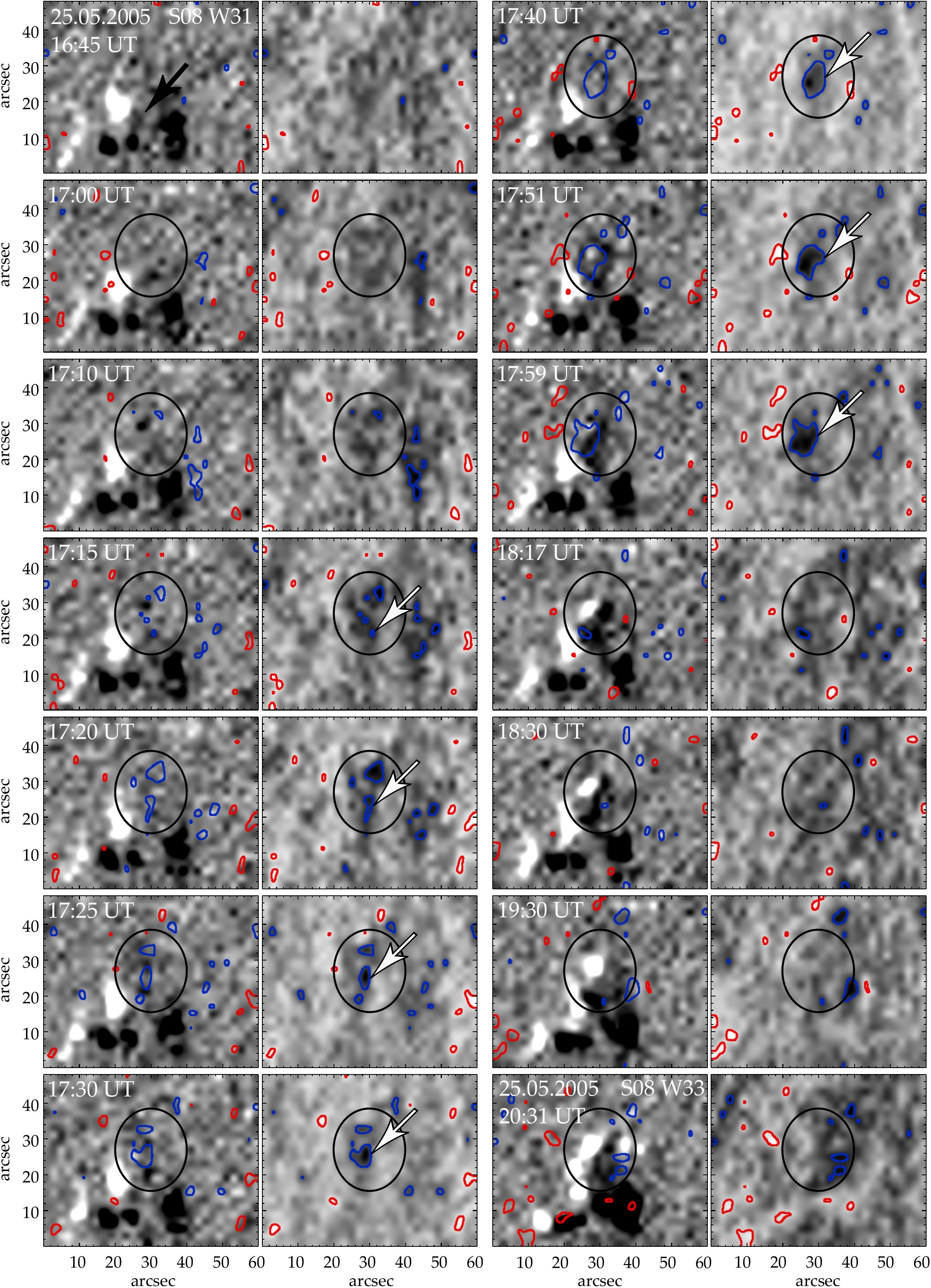}}
\caption{Emergence of the active region NOAA 10768 on 2005 May 25. Notations 
are the same as in Figure \ref{fig:fig2}. The region of magnetic flux emergence 
is marked by ellipse. See the comments in the text.}
\label{fig:fig5}
\end{figure*}

\textit{Initial stage of the magnetic flux emergence.} Before the start of the 
magnetic flux emergence of NOAA 10768, there were several pre-existing magnetic 
elements of a scale of an ephemeral active region near this place, which 
appeared on May 25, $\sim$02:00 UT (Figure \ref{fig:fig5}, marked by a black 
arrow on the magnetogram at 12:10 UT). NOAA 10768 begins to emerge on May 25, 
17:42 UT, as a negative magnetic field (leading polarity). The shape and size of 
the $-$60\,G isoline show a drastic evolution. About 16 minutes later, a 
positive field (following polarity) appears. The time lag in the appearance and 
the size asymmetry of the both polarities may attribute to the horizontal 
magnetic field at the apex of the emerging $\Omega$-loop.

It is clearly seen from Figure \ref{fig:fig5} that the negative Doppler velocity 
begins to enhance before the appearance of the line of sight magnetic field in 
the photosphere. Upflow inside the $-$500\,m\,s$^{-1}$ isoline arises from 17:15 
UT (Figure \ref{fig:fig5}, marked by white arrows). The enhanced negative 
Doppler velocity is localized, mainly on the polarity inversion line between the 
negative and positive polarities. The region of the enhanced plasma upflow 
disappears during the data gap from 18:03 UT to 18:17 UT, approximately 1 hr 
after the beginning of emergence.

\begin{figure*}
\centerline{\includegraphics[width=0.45\textwidth]{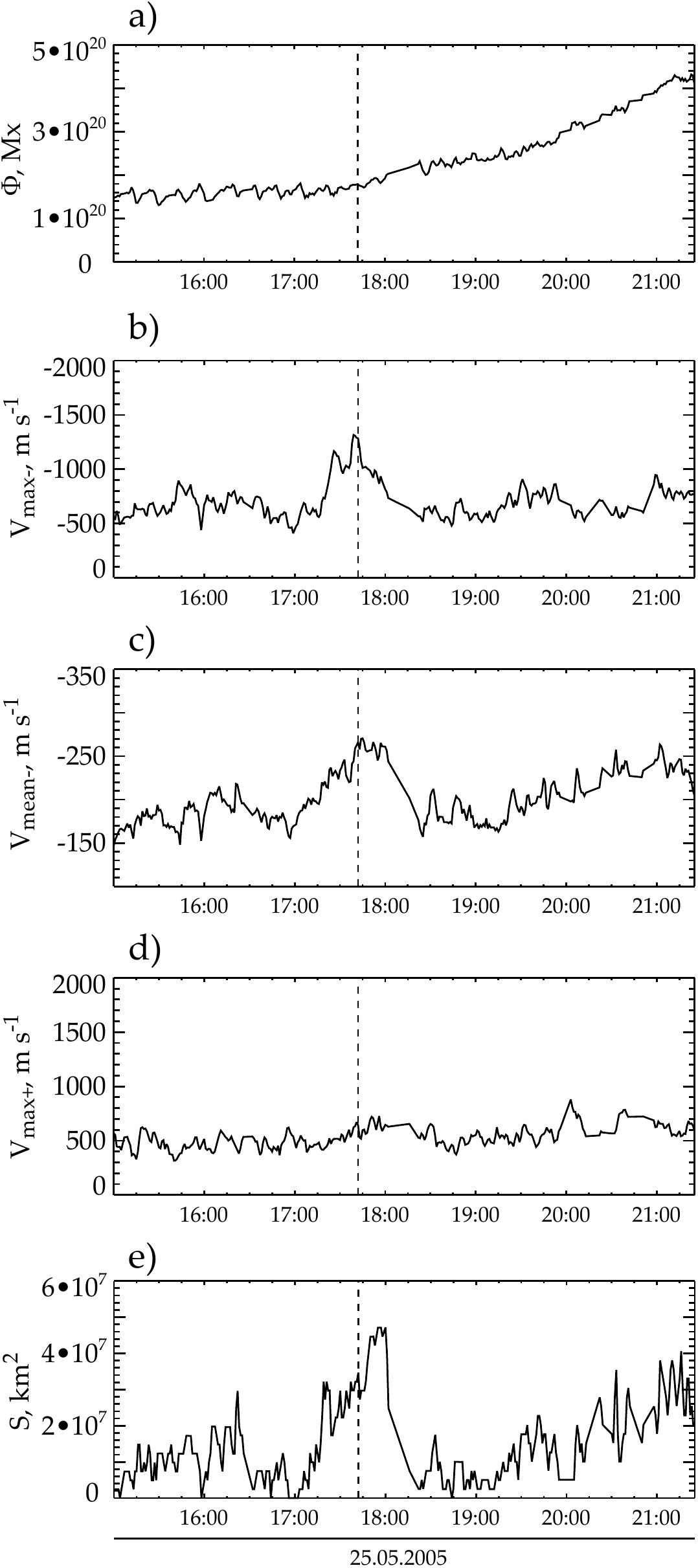}}
\caption{The active region NOAA 10768. Notations are the same as in 
Figure \ref{fig:fig3}.}
\label{fig:fig6}
\end{figure*}

Figure \ref{fig:fig6} shows the time variation for parameters of the magnetic 
field and plasma flows during the emergence of NOAA 10768. The growth of the 
total magnetic flux begins at 17:42 UT (Figure \ref{fig:fig6}\,a). However, the 
enhancement of maximum negative Doppler velocity appears 27 minutes earlier at 
17:15 UT (Figure \ref{fig:fig6}\,b). Plasma upflow velocity reaches the peak 
value of $-$1320\,m\,s$^{-1}$ at 17:40 UT, which is 25 minutes after the 
velocity enhancement starts and 2 minutes before the magnetic field appears in 
the photosphere. After that, the maximum negative Doppler velocity reduces. The 
mean negative Doppler velocity increases approximately 1.5 times (Figure 
\ref{fig:fig6}\,c). The mean and maximum negative Doppler velocities reach their 
peak values about at the same time. The maximum positive Doppler velocity does 
not exceed 730\,m\,s$^{-1}$ during this period (Figure \ref{fig:fig6}\,d). The 
region with enhanced negative Doppler velocity remains for about 1 hr, with the 
maximum area inside the $-$500\,m\,s$^{-1}$ isoline reaching 
$\sim$4.7$\times10^{7}$\,km$^{2}$ (Figure \ref{fig:fig6}\,e).

\subsection{Comparison to a large active region}\label{sec:ar_upflow}

\floattable
\begin{deluxetable}{cccccc}
\tablecaption{Main characteristics of active regions: $S_{\rm spot}$, the 
sunspot area, and $\Phi_{\rm max}$, the total unsigned magnetic flux, at the maximum 
development; \textit{d$\Phi$/dt}, the total unsigned magnetic flux growth rate, 
and $H_{\rm max}$, maximum magnetic field strength, in the first 12 hr of 
development of the active regions \label{tab:ar_main2}}
\tablecolumns{6}
\tablenum{2}
\tablewidth{0pt}
\tablehead{\colhead{Active Region} & \colhead{$S_{\rm spot}$} & \colhead{$\Phi_{\rm max}$} & \colhead{d$\Phi$/dt} & \colhead{$H_{\rm max}$} \\
\colhead{} & \colhead{(MSH)} & \colhead{(Mx)} & \colhead{(Mx\,hr$^{-1}$)} & \colhead{(G)}}
\startdata
 9021 &   19 &  $2.9\times10^{21}$ & $0.5\times10^{20}$ &  990 \\
10768 &   14 &  $2.2\times10^{21}$ &   $1\times10^{20}$ & 1100 \\
\hline
10488 & 1725 & $>5.9\times10^{22}$ &   $4\times10^{20}$ & 1630 \\
\enddata
\end{deluxetable}

\floattable
\begin{deluxetable}{cccccccc}
\tablecaption{Characteristics of velocity structures: $V_{\rm max+}$ and $V_{\rm 
max-}$ are the maximum positive and negative Doppler velocities during 
enhancement of negative Doppler velocity; $S_{\rm vel}$, the maximum area of 
velocity structure inside the $-$500\,m\,s$^{-1}$ isoline; $t_{\rm before}$, the 
time lag between the appearance of the enhanced upflow structure and the initial 
magnetic flux emergence; $t_{\rm max}$, the time lag between the enhanced upflow 
appearance and the peak upflow velocity; and $t_{\rm life}$, the duration when 
the enhanced upflow structure exists \label{tab:ar_main3}}
\tablecolumns{8}
\tablenum{3}
\tablewidth{0pt}
\tablehead{\colhead{Active Region} & \colhead{$V_{\rm max+}$} & \colhead{$V_{\rm max-}$} & \colhead{$S_{\rm vel}$} & \colhead{$t_{\rm before}$} & \colhead{$t_{\rm max}$} & \colhead{$t_{\rm life}$} \\
\colhead{} & \colhead{(m\,s$^{-1}$)} & \colhead{(m\,s$^{-1}$)} & \colhead{(km$^{2}$)} & \colhead{(minutes)} & \colhead{(minutes)} & \colhead{(hr)}}
\startdata
 9021 & 720 & $-$1650 & $\sim5.7\times10^{7}$ & 27 & 40 & $\sim$2 \\
10768 & 730 & $-$1320 & $\sim4.7\times10^{7}$ & 27 & 25 & $\sim$1 \\
\hline
10488 & 840 & $-$1680 & $\sim5.4\times10^{7}$ & 12 & 20 & $\sim$2 \\
\enddata
\end{deluxetable}

In the previous subsections, we investigated the detailed evolution of magnetic 
and velocity fields in the emerging sites of the two small-scale active regions 
NOAA 9021 and 10768. Here we compare the observational results with those of a 
large active region NOAA 10488 \citep{grigorev07}.

Table \ref{tab:ar_main2} compares the parameters that characterize the entire 
active regions. The total sunspot area of NOAA 10488, 1725 MSH, is by far larger 
than those of the active regions NOAA 9021 and 10768, which are 19 and 14 MSH, 
respectively. Similarly, the total magnetic flux, magnetic flux growth rate, and 
maximum magnetic strength show great differences between the compared active 
regions. From these parameters, NOAA 10488 can be regarded as a large active 
region and NOAA 9021 and 10768 as small active regions (see, e.g., 
\citealt{zwaan87}).

On the other hand, most of the parameters that describe velocity structures at 
the initial stage of the magnetic flux emergence in Table \ref{tab:ar_main3} do 
not show a clear distinction between the large and small active regions. For 
example, the area of the enhanced upflow is $\sim$5.4$\times10^{7}$\,km$^{2}$, 
which is equivalent to a circular region with a diameter of $\sim$8\,Mm for NOAA 
10488 and $\sim$5.7$\times10^{7}$\,km$^{2}$ and 
$\sim$4.7$\times10^{7}$\,km$^{2}$ for the two small active regions NOAA 9021 and 
10768. Other parameters are also comparable, except for the time lag between the 
upflow start and the initial flux appearance.

Similar parameters of enhanced upflows in the earliest phase of magnetic flux 
emergence in active regions of different spatial scales may point to the 
emergence of elementary magnetic flux with similar properties. For making firm 
conclusions, we may need a statistical analysis. We found the following general 
trends: (1) the enhanced upflow precedes the magnetic flux appearance, (2) the 
upflow velocity attains its peak value about the same time or after the magnetic 
flux emergence starts, and (3) the enhanced upflow area becomes largest after 
these timings (emergence start and upflow peak).

\section{Physical Mechanism of the Plasma Upflow} \label{sec:phys}

We analyze the results of numerical simulations conducted by \cite{toriumi11} to 
investigate the driving mechanism of the photospheric plasma upflows at the 
initial stage of formation of the active regions that we studied in Section 
\ref{sec:ar}. This model is based on solving three-dimensional ideal MHD 
equations. The physical values are normalized by the photospheric pressure scale 
height $H_{0}$=170\,km, the sound speed $C_{\rm s0}=6.8\ {\rm km\ s^{-1}}$, the 
sound crossing time $\tau_{0}=H_{0}$/$C_{s0}$=25\,s, and the magnetic field 
strength $B_{0}$=250\,G.

Calculations are performed in a rectangular domain that has the size 
($-$120,$-$120,$-$20)$\leq$($x/H_{0}$,$y/H_{0}$,$z/H_{0}$)$\leq$(120,120,150) or 
40.8\,Mm$\times$40.8\,Mm$\times$28.9\,Mm with the grid number of 
256$\times$256$\times$256. The stratification of simulation box includes three 
regions: an adiabatically stratified convective zone ($z/H_{0}<$0), a low 
temperature isothermal photosphere/chromosphere (0$\leq z/H_{0}<$10), and a high 
temperature isothermal corona ($z/H_{0}\geq$20). The horizontal magnetic flux 
tube is initially located in the convection zone at $z$=$-10H_{0}=-1.7$\,Mm. It 
has the radius $R_{\rm tube}$=2.5$H_{0}$=425\,km, the axial magnetic field 
strength $B_{\rm tube}$=15$B_{0}$=3750\,G, and the twist parameter 
$q$=0.2/$H_{0}$=1.1$\times10^{-3}$\,km$^{-1}$. The total magnetic flux of flux 
tube is equal to $\Phi$=2.1$\times10^{19}$\,Mx. In this section, we compare the 
simulation results with the elementary emerging loops in the active regions we 
analyzed in Section \ref{sec:ar}, not with the active regions themselves.

\begin{figure*}
\centerline{\includegraphics[width=0.45\textwidth]{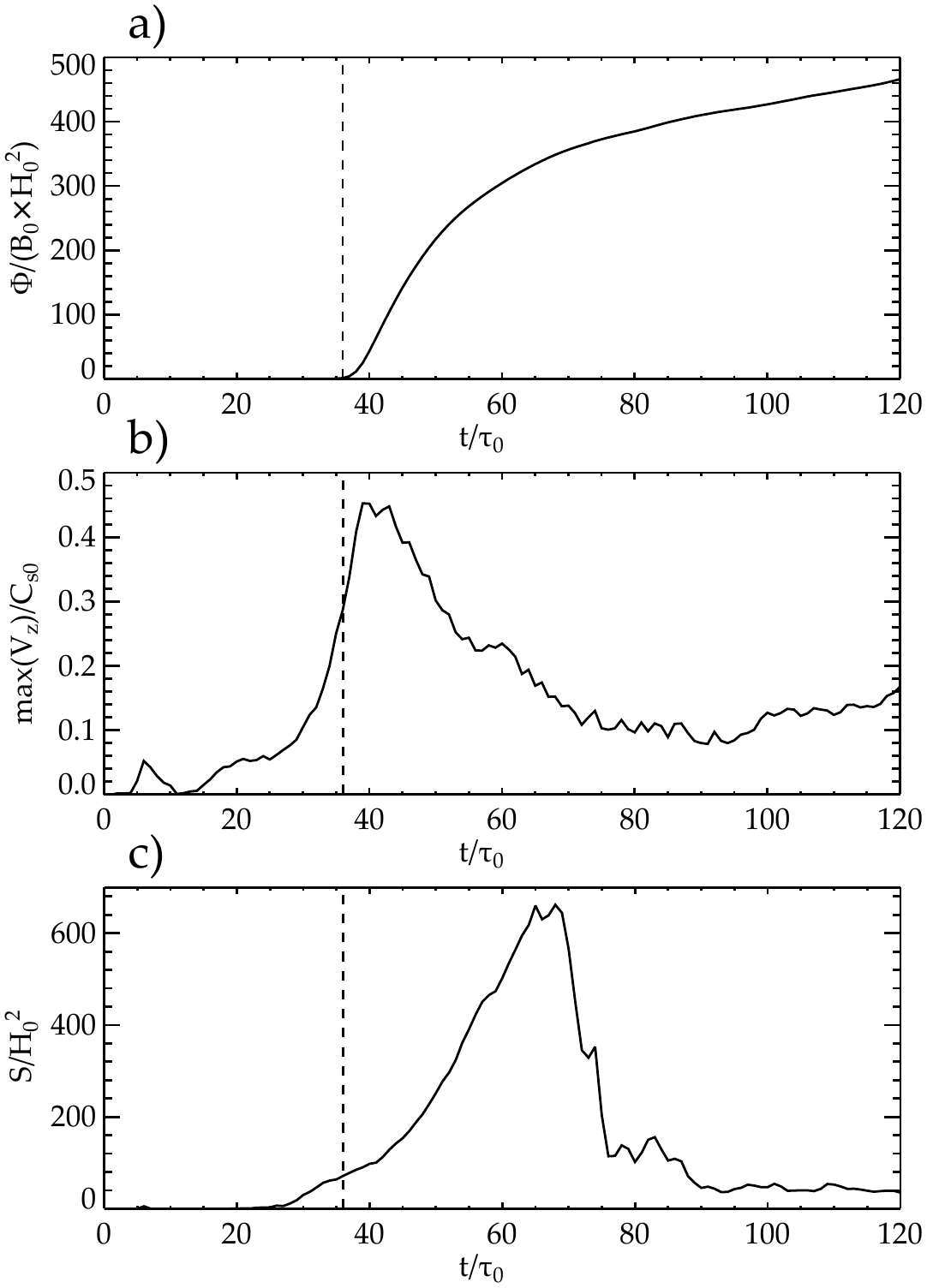}}
\caption{Time variation in (a) the total unsigned magnetic flux, (b) the 
maximum upflow velocity, and (c) the area of plasma upflow inside the 
$V_{z}$/$C_{s0}$=0.05 isoline on the solar surface at the $z/H_{0}$=0 height. 
The vertical dashed line marks the beginning of the magnetic flux emergence.}
\label{fig:fig7}
\end{figure*}

In Figure \ref{fig:fig7}, we plot the time variation of the total magnetic flux, 
the maximum velocity of the plasma upflow, and the upflow area inside the 
$V_{z}$/$C_{s0}$=0.05 isoline on the solar surface at $z/H_{0}$=0. In the 
simulation a positive vertical velocity ($V_{z}>$0) corresponds to the upward 
plasma flow, which is shown as a negative Doppler velocity in the observations. 
The magnetic flux emergence on the surface begins at $t/\tau_{0}\approx$36 
(Figure \ref{fig:fig7}\,a). However, the strong upflow characterized by the 
$V_{z}$/$C_{s0}$=0.05 isoline appears earlier at $t/\tau_{0}\approx$25 (Figure 
\ref{fig:fig7}\,b). The upflow velocity reaches its maximal value of 
$V_{z}\approx0.45\times C_{\rm s0}\approx3$\,km\,s$^{-1}$ at 
$t/\tau_{0}\approx$40 and then slowly decreases. The area of the upflow reaches 
its peak value of $S\approx 660 \times H_{0}^{2}\approx 1.9 
\times10^{7}$\,km$^{2}$ at $t/\tau_{0}\approx$68 (i.e., after the upflow 
velocity attains its maximum), and then vanishes rapidly (Figure 
\ref{fig:fig7}\,c). The time evolution of velocity and area of plasma upflow on 
the solar surface obtained here agrees quite well with the observational results 
summarized in Section \ref{sec:ar_upflow} (also compare Figure \ref{fig:fig7} 
with Figures \ref{fig:fig3} and \ref{fig:fig6}). The striking consistency 
between the observation and the simulation suggests a common physical background 
for the enhanced plasma upflow.

\begin{figure*}
\centerline{\includegraphics[width=0.75\textwidth]{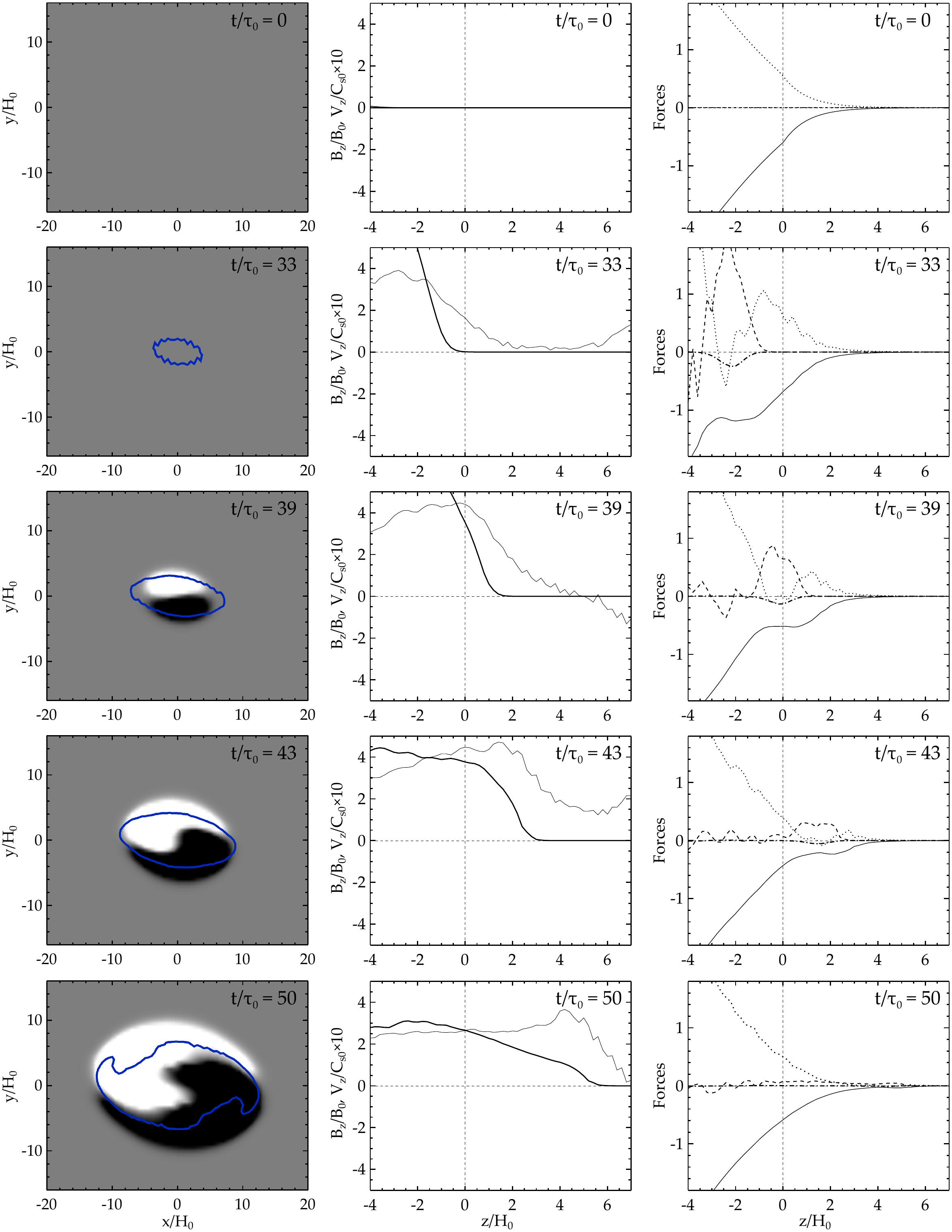}} 
\caption{Left column: the surface magnetograms with the superimposed isoline of 
the $V_{z}$/$C_{s0}$=0.1 plasma upflow velocity at different instants. Middle 
column: the distribution of the magnetic field strength (thick line) and of the 
velocity (thin line) along the vertical axis $z$ at $x/H_{0}$=$y/H_{0}$=0. Right 
column: the distribution of the vertical component of the forces along the axis 
$z$ at $x/H_{0}$=$y/H_{0}$=0: the gas pressure gradient (dotted line), the 
magnetic pressure gradient (dashed line), the magnetic tension (dashed-dotted 
line), and the gravity (thin solid line). The vertical dashed line on the plots 
marks the solar surface $z/H_{0}$=0.}
\label{fig:fig8}
\end{figure*}

To understand the mechanism of the upflow during the magnetic flux emergence, we 
measure the forces that act on the plasma and investigate the dynamics. Figure 
\ref{fig:fig8} shows the surface magnetogram with the $V_{z}$/$C_{s0}$=0.1 
isoline of plasma upflow velocity (left column), the distributions of the 
magnetic field strength and the vertical velocity along the $z$-axis at 
$x/H_{0}$=$y/H_{0}$=0 (middle column), and the distributions of the vertical 
component of the forces along the same axis (right column) for different 
instants. The left and middle columns at $t/\tau_{0}$=33 show that the increase 
in the upflow velocity on the surface arises before the emerging magnetic flux 
appears. The right column demonstrates that this occurs due to an increase in 
the gas pressure gradient. Earlier, \cite{cheung10} and \cite{toriumi13a} found 
that the increase in the gas pressure gradient at the apex of the emerging 
magnetic flux drives the plasma outflows at the solar surface. The increase in 
the gas pressure gradient happens due to the compression of the plasma by the 
emerging flux \citep{toriumi13a}. However, when the magnetic flux reaches the 
surface ($z/H_{0}$=0), the gas pressure gradient is suppressed and the magnetic 
pressure gradient becomes dominant (right column at $t/\tau_{0}$=39). Therefore, 
at this stage, the primary force that drives the upflow is the magnetic pressure 
gradient instead of the gas pressure gradient. In the later stage, the upflow 
velocity decreases because the emerging magnetic flux expands into the solar 
atmosphere and the magnetic pressure gradient becomes less pronounced 
($t/\tau_{0}$=43 and 50).

Therefore, we can conclude that there exists two distinct driving forces for the 
plasma upflows, depending on the evolution stage. Until just before the magnetic 
flux reaches the photosphere, the upflow is caused by the gas pressure gradient, 
while after the magnetic flux appears in the photosphere, the magnetic pressure 
gradient plays a key role.

\section{Discussion} \label{sec:disc}

In this study, we analyzed the plasma flow structures that accompany the 
magnetic flux emergence events in the photosphere, especially of small active 
regions with the total unsigned magnetic flux of $\sim10^{21}$\,Mx. The detected 
enhanced upflow region that appears before the magnetic flux emergence has a 
negative Doppler velocity of up to $\sim-$1600\,m\,s$^{-1}$, an area of 
$\sim$5$\times10^{7}$\,km$^{2}$ (or 8\,Mm in diameter), and continues for 
1\,--\,2 hr. Such a flow structure is easily distinguished from the regular 
granules and supergranules in the quiet Sun. For example, the observed upflow 
velocity (1600\,m\,s$^{-1}$) is much faster than the typical upflow velocity of 
supergranules (300\,m\,s$^{-1}$), while the size scale of the enhanced upflow 
(8\,Mm) is much larger than that of the granules (0.5\,--\,2\,Mm): parameters of 
granules and supergranules are taken from \cite{rieutord10}. Also, the lifetime 
of the velocity structure with enhanced upflows (1\,--\,2 hr) is not relevant to 
those of the regular quiet-Sun granules (10 minutes) and supergranules (1\,--\,2 
days). Moreover, the observed upflows may not be the large granular convection 
cells, expanded due to the emerging magnetic flux, since such cells may appear 
over the whole period of the magnetic flux emergence, while the presented 
upflows are seen only at the initial stage of active region development.

One of the most interesting results from the observation is that active regions 
of different sizes (more than one order of the magnetic flux magnitude) show 
similar parameters for the enhanced upflow such as maximum velocity, area, and 
lifetime (see Section \ref{sec:ar_upflow}). This may point to the possibility 
that although the large-scale structures of the active regions are different, 
the elementary magnetic flux tubes that comprise the entire emerging flux have 
similar physical parameters for various events (tube size, magnetic field 
strength, rise velocity, etc.). Perhaps the branching of the original subsurface 
flux system into elementary tubes is due to local convection, and thus the 
elementary tubes have similar properties regardless of the size of active 
regions.

It has been widely considered that magnetic buoyancy is the probable driver of 
the rising magnetic flux in the solar interior (\citealt{parker55} and others). 
Recently \cite{khlystova13b} found from the statistical analysis that the 
magnetic flux growth rate is proportional to the square of the magnetic field 
strength on the initial stage of the active region appearance. Since the 
magnetic buoyancy has a quadratic dependence on the magnetic field strength, 
this result shows that the magnetic buoyancy has an important role in the 
magnetic flux emergence on the solar surface. Also, the numerical simulations 
show influence of the convective flows on the magnetic flux emergence inside the 
convective zone \citep{fan03,weber11,weber13,jouve13} and near the solar surface 
\citep{cheung07,cheung08,martinez08,yelles09,tortosa09,fang10,stein11,bushby12}. 
Radiative MHD simulations of the formation of active regions by \cite{cheung10} 
and \cite{rempel14} demonstrate the formation of undulated structure of field 
lines through the interaction of the emerging magnetic flux with the small-scale 
granular convection. Simulations by \cite{fang12} and \cite{stein12} show that 
the large-scale convective motions concentrate magnetic fields and play an 
essential role in the formation of pores and sunspots. Therefore, it is possible 
that the enhanced upflow structures in emerging magnetic flux regions detected 
in this paper are coupled with large-scale convective motions. Still, the upflow 
patterns seen at the photosphere are to a large extent different from those of 
regular granules and supergranules.

\section{Conclusions} \label{sec:conc}

We found a strong plasma upflow during the emergence of two small active regions 
NOAA 9021 and NOAA 10768 in the solar photosphere. The values of maximum 
negative Doppler velocity reach $-$1650\,m\,s$^{-1}$ and $-$1320\,m\,s$^{-1}$, 
respectively. The enhanced upflow regions have sizes of 
$\sim$5.7$\times10^{7}$\,km$^{2}$ and $\sim$4.7$\times10^{7}$\,km$^{2}$, which 
correspond to $\sim$8\,Mm in diameter. The increase of the plasma upflow 
velocity begins 27 minutes before the appearance of the magnetic flux in the 
photosphere. The lifetime of enhanced upflow is $\sim$2 and $\sim$1 hr. The 
observed flows are similar to that which appeared during the emergence of the 
large active region NOAA 10488 \citep{grigorev07}.

We used numerical simulations by \cite{toriumi11} to investigate the causes of 
the observed plasma flows. The change of plasma upflow velocity during the 
magnetic flux emergence at the solar surface can be explain by the action of two 
forces -- namely, the gas pressure gradient and the magnetic pressure gradient. 
The increase in the plasma upflow velocity before the emerging magnetic flux 
occurs due to an increase of the gas pressure gradient. When the magnetic flux 
penetrates into the solar photosphere, the plasma upflow velocity continues to 
enhance due to an increase of the magnetic pressure gradient. After the decrease 
in the magnetic pressure gradient due to the expansion of the emerging magnetic 
flux into the solar atmosphere, the upflow velocity on the surface began to 
decrease.

The similar parameters of enhanced upflows at the appearance of small and large 
active regions may point to the possibility that the elementary magnetic flux 
tubes at the very beginning of emergence have similar initial properties, 
irrespective of scales of active regions. It is possible that there is an 
additional contribution of the convective upflows to rising of the magnetic 
fluxes to the solar surface.

\acknowledgments

We thank the anonymous referee for improving the manuscript. This work was 
supported by JSPS and RFBR under the Japan-Russia Research Cooperative Program. 
A.K. thanks the support of RFBR grant 16-52-50077 and the ISTP SB RAS project 
II.16.3.1. S.T. is supported by JSPS KAKENHI Grant Numbers JP16K17671 and 
JP15H05814. We used data obtained by the \textit{SOHO/}MDI instrument. 
\textit{SOHO} is a project of international cooperation between ESA and NASA.

\bibliography{ms}
\bibliographystyle{aasjournal}

\newpage

\end{document}